%% file: Multiuser_AO_Rev_v1.tex
\documentclass[twocolumn,conference]{IEEEtran}
\PassOptionsToPackage{ruled}{algorithm2e}
\usepackage[T1]{fontenc}
\usepackage[latin9]{inputenc}
\usepackage{color}
\usepackage{algorithm2e}
\usepackage{amsmath}
\usepackage{amssymb}

\makeatletter

\DeclareMathOperator{\maximize}{maximize}
\DeclareMathOperator{\minimize}{minimize}
\DeclareMathOperator{\st}{subject~to}
\DeclareMathOperator{\diag}{diag}

\DeclareMathOperator{\tr}{Tr}
\DeclareMathOperator{\rank}{rank}

\newcommand{\herm}{^{{\dagger}}}

\usepackage{cite}
\usepackage{acronym}
\AtBeginDocument{\input{acro.tex}}

\usepackage{pgfplots}
\usepackage{grffile}
\pgfplotsset{compat=newest}
\usetikzlibrary{plotmarks}
\usepgfplotslibrary{patchplots}
\usepackage{amsmath}

\definecolor{mycolor1}{rgb}{1.00000,0.00000,1.00000}%

\pgfplotsset{compat=newest}
\usetikzlibrary{plotmarks}
\usepgfplotslibrary{patchplots}

\tikzset{every node/.style={font=\small}}
\tikzset{every pin/.style={fill=white,font=\small}}
\tikzset{every pin edge/.style={<-,>=stealth,black,thick}}
\pgfplotsset{grid style={dotted,gray}}
\pgfplotsset{every axis/.style={inner sep=2pt}}
\pgfplotsset{legend style={font=\small}}
\newlength\figurewidth
\newlength\figureheight

\definecolor{mycolor1}{rgb}{1.00000,0.00000,1.00000}%

\makeatother

\begin{document}
\title{On the Achievable Sum-rate of the RIS-aided MIMO Broadcast Channel}
\author{\IEEEauthorblockN{Nemanja~Stefan~Perovi\'c\IEEEauthorrefmark{1}, Le-Nam Tran\IEEEauthorrefmark{1},
Marco~Di~Renzo\IEEEauthorrefmark{2}, and Mark~F.~Flanagan\IEEEauthorrefmark{1}}\IEEEauthorblockA{\IEEEauthorrefmark{1}School of Electrical and Electronic Engineering,
University College Dublin\\
Belfield, Dublin 4, D04 V1W8, Ireland\\
Email: nemanja.stefan.perovic@ucd.ie, nam.tran@ucd.ie, mark.flanagan@ieee.org}\IEEEauthorblockA{\IEEEauthorrefmark{2}Université Paris-Saclay, CNRS, CentraleSupélec,
Laboratoire des Signaux et Systèmes\\
3 rue Joliot Curie, 91192, Gif-sur-Yvette, France\\
Email: marco.di-renzo@universite-paris-saclay.fr}}
\IEEEspecialpapernotice{Invited Paper}
\maketitle
\begin{abstract}
Reconfigurable intelligent surfaces (RISs)\acused{RIS} represent
a new technology that can shape the radio wave propagation and thus
offers a great variety of possible performance and implementation
gains. Motivated by this, we investigate the achievable sum-rate optimization
in a \ac{BC} that is equipped with an \ac{RIS}. We exploit the well-known
duality between the Gaussian \ac{MIMO} \ac{BC} and \ac{MAC} to
derive an \ac{AO} algorithm which jointly optimizes the users' covariance
matrices and the RIS phase shifts in the dual MAC. The optimal users'
covariance matrices are obtained by a dual decomposition method in
which each iteration is solved in closed-form. The optimal RIS phase
shifts are also computed using a derived closed-form expression. Furthermore,
we present a computational complexity analysis for the proposed AO
algorithm. Simulation results show that the proposed \ac{AO} algorithm
can provide significant achievable sum-rate gains in a \ac{BC}.\acresetall{}
\end{abstract}


\section{Introduction}

\textcolor{black}{\bstctlcite{BSTcontrol}}The need to satisfy constantly
increasing data rate demands in wireless communication networks motivates
the development of new technology solutions such as \acp{RIS}. An
\ac{RIS} is a thin metasurface that consists of a large number of
small, low-cost, and passive elements. Since each of these elements
can reflect the incident signal with an adjustable phase shift, an
RIS can effectively shape the propagation of the impinging wave \cite{di2019smart,di2020smart}.
Therefore, the introduction of \acp{RIS} offers a wide variety of
possible implementation gains and potentially presents a new milestone
in wireless communications.

In order to fully exploit the gains that arise from the use of \acp{RIS},
we need to obtain a deep understanding of different aspects of RIS-assisted
wireless communication systems. Probably the most important aspect
concerns the optimal design of the RIS phase shifts, so that the incoming
radio wave is altered in a way that maximizes the aforementioned gains.
In this regard, the development of algorithms for the achievable rate
optimization is of particular interest for \mbox{\ac{RIS}-aided}
communications. A significant body of research work in this area concentrates
on the achievable rate optimization for point-to-point communications
with continuous \cite{perovic2020achievable,zhang2019capacity,perovic2019channel}
and discrete \cite{perovic2020optimization} signaling. Another equally
important part of this research work has been dedicated to the achievable
sum-rate optimization for multi-user systems. In \cite{zhang2020intelligent},
the authors studied the capacity/achievable rate region for the \ac{MAC}
and for the \ac{BC} by using the well-known \ac{BC}-\ac{MAC} duality;
however, the analysis was limited to single-antenna user terminals
and a single-antenna \ac{BS}. Methods for optimization of the achievable
sum-rate for multi-user \ac{MISO} communication systems equipped
with \acp{RIS} were introduced in \cite{guo2019weighted,kammoun2020asymptotic}.
The use of an RIS in multi-cell \ac{MIMO} systems was investigated
in \cite{pan2020multicell}, where the aim was to improve downlink
transmission to cell-edge users by employing an RIS which increases
the weighted sum-rate in the considered communication system. An extension
of this work to the case of an RIS-aided MIMO system performing simultaneous
wireless information and power transfer was presented in \cite{pan2020intelligent}.

Against this background, the contributions of this paper are listed
as follows:
\begin{itemize}
\item We exploit the Gaussian MIMO \ac{BC}-\ac{MAC} duality to maximize
the achievable sum-rate of a MIMO system equipped with an RIS communicating
over a BC, and formulate a joint optimization problem of the users'
covariance matrices and the RIS elements' phase shifts. To solve this
problem, we propose an iterative algorithm which operates in the dual
MAC, and optimizes the users' covariance matrices and the RIS elements'
phase shifts in an alternating manner. The optimal users' covariance
matrices are obtained by a dual decomposition method, while the optimal
RIS phase shifts are computed by a derived closed-form expression.
\item For the proposed \ac{AO} algorithm, we derive an expression for the
computational complexity in terms of the number of complex multiplications.
\item Simulation results show that the AO can provide significant achievable
sum-rate gains. These gains increase with the number of users and
the number of transmit antennas, especially when the direct links
are present in the BC.
\end{itemize}
\textit{Notation}: Bold lower and upper case letters represent vectors
and matrices, respectively. $\mathbb{C}^{m\times n}$ denotes the
space of $m\times n$ complex matrices. $\mathbf{H}^{T}$ and $\mathbf{H}\herm$
denote the transpose and Hermitian transpose of $\mathbf{H}$, respectively;
$|\mathbf{H}|$ is the determinant of $\mathbf{H}$. $\tr(\mathbf{H})$
stands for the trace of $\mathbf{H}$ and $\rank(\mathbf{H})$ denotes
the rank of $\mathbf{H}$. $\log_{2}(\cdot)$ is the binary logarithm,
$\ln(\cdot)$ is the natural logarithm and $(x)_{+}$ denotes $\max(0,x)$.
$\mathbb{E}\bigl\{\cdot\bigr\}$ stands for the expectation operator
and $\left(\cdot\right)^{\ast}$ denotes the complex conjugate.\textcolor{black}{{}
The notation $\mathbf{A}\succeq(\succ)\mathbf{B}$ means that $\mathbf{A}-\mathbf{B}$
is positive semidefinite (definite).} $\mathbf{I}$ represents an
identity matrix whose size should be clear from the context. For a
vector~$\mathbf{x}$, $\diag(\mathbf{x})$ denotes a diagonal matrix
with the elements of $\mathbf{x}$ on the diagonal. $\mathcal{CN}(\mu,\sigma^{2}$)
denotes a circularly symmetric complex Gaussian random variable of
mean $\mu$ and variance~$\sigma^{2}$.

\section{System Model}

We consider a \ac{BC} in which one \ac{BS} simultaneously serves
$K$ users. Both the \ac{BS} and the users are equipped with multiple
antennas, such that the \ac{BS} and the \emph{k}-th user have $N_{t}$
and $n_{k}$ antennas, respectively. The \ac{BS} antennas are placed
in a \ac{ULA} with inter-antenna separation $s_{t}$. In a similar
manner, all the antennas of a single user are placed in a \ac{ULA}
with inter-antenna separation $s_{r}$. In order to improve the system
performance, an \ac{RIS} is also present in the considered communication
environment. It consists of $N_{\mathrm{ris}}$ reflecting elements
which are placed in a \ac{URA}, so that the separation between the
centers of adjacent \ac{RIS} elements in both dimensions is $s_{\mathrm{ris}}$.

The received signal at the \emph{k}-th user is given by 
\begin{align}
\mathbf{y}_{k} & =\mathbf{H}_{k}\mathbf{x}_{k}+\sum\nolimits _{j=1,j\neq k}^{K}\mathbf{H}_{k}\mathbf{x}_{j}+\mathbf{n}_{k}\label{eq:sigmod:gen}
\end{align}
where $\mathbf{H}_{k}\in\mathbb{C}^{n_{k}\times N_{t}}$ is the channel
matrix for the \emph{k}-th user, $\mathbf{x}_{k}\in\mathbb{C}^{N_{t}\times1}$
is the transmitted signal intended for the \emph{k}-th user, and $\mathbf{x}_{j}\in\mathbb{C}^{N_{t}\times1}$
for $j\neq k$ are the transmitted signals intended for other users,
which act as interference for the detection of $\mathbf{x}_{k}$.
The noise vector $\mathbf{n}_{k}\in\mathbb{C}^{n_{k}\times1}$ consists
of \ac{iid} elements that are distributed according to $\mathcal{CN}(0,N_{0})$,
where $N_{0}$ is the noise variance.

Due to the presence of the RIS, the channel matrix $\mathbf{H}_{k}$
can be expressed as
\begin{equation}
\mathbf{H}_{k}=\mathbf{D}_{k}+\mathbf{G}_{k}\mathbf{F}(\boldsymbol{\theta})\mathbf{U}\label{eq:Hk_equ}
\end{equation}
where $\mathbf{D}_{k}\in\mathbb{C}^{n_{k}\times N_{t}}$ is the direct
link channel matrix between the \ac{BS} and the \emph{k}-th user,
$\mathbf{U}\in\mathbb{C}^{N_{\mathrm{ris}}\times N_{t}}$ is the channel
matrix between the BS and the RIS, and $\mathbf{G}_{k}\in\mathbb{C}^{n_{k}\times N_{\mathrm{ris}}}$
is the channel matrix between the RIS and the \emph{k}-th user. Signal
reflection from the RIS is modeled by $\mathbf{F}(\boldsymbol{\theta})=\mathrm{diag}(\boldsymbol{\theta})\in\mathbb{C}^{N_{\mathrm{ris}}\times N_{\mathrm{ris}}}$,
where $\boldsymbol{\theta}=[\theta_{1},\theta_{2},\ldots,\theta_{N_{\mathrm{ris}}}]^{T}\in\mathbb{C}^{N_{\mathrm{ris}}\times1}$.
We assume that the signal reflection from any RIS element is ideal
(i.e., without any power loss) and therefore we may write $\theta_{l}=e^{j\phi_{l}}$
for $l=1,2,\ldots,N_{\mathrm{ris}}$, where $\phi_{l}$ is the phase
shift induced by the $l$-th RIS element. Equivalently, this can be
written as
\begin{equation}
\left|\theta_{l}\right|=1,\quad l=1,2,\ldots,N_{\mathrm{ris}}.\label{eq:RIS_elem_cons}
\end{equation}

\section{Problem Formulation}

In this paper, we are interested in maximizing the achievable sum-rate
of the considered RIS-assisted wireless communication system. To accomplish
this, we exploit the fact that the achievable rate region of a Gaussian
MIMO BC can be achieved by \ac{DPC} \cite{Weingarten:CapacityRegion:MU_MIMO:2006}.
\ac{DPC} enables us to perfectly eliminate the interference term
$\sum_{j<k}\mathbf{H}_{k}\mathbf{x}_{j}$ for the \emph{k}-th user,
assuming that the BS has full (non-causal) knowledge of this interference
term. In this regard, the ordering of the users clearly matters. Let
$\pi$ be an ordering of users, i.e., a permutation of the set $\{1,2,\ldots,K\}$.
Then for this ordering, the achievable rate for the \emph{k}-th user
can be computed as \cite[Eq. (3)]{Vishwanath:duality_achievable:2003}{\small
\begin{equation}
R_{\pi(k)}=\log_{2}\frac{\Bigl|\mathbf{I}+\mathbf{H}_{\pi(k)}\bigl(\sum_{j\geq k}\mathbf{S}_{\pi(j)}\bigr)\mathbf{H}_{\pi(k)}\herm\Bigr|}{\Bigl|\mathbf{I}+\mathbf{H}_{\pi(k)}\bigl(\sum_{j>k}\mathbf{S}_{\pi(j)}\bigr)\mathbf{H}_{\pi(k)}\herm\Bigr|},k=1,2,\ldots,K
\end{equation}
}where $\mathbf{S}_{k}=\mathbb{E}\bigl\{\mathbf{x}_{k}\mathbf{x}_{k}\herm\bigr\}\succeq\mathbf{0}$
is the input covariance matrix of user $k$. In this paper, we consider
a sum power constraint at the BS, i.e.,{\small
\begin{equation}
\sum\nolimits _{k=1}^{K}\tr\bigl(\mathbf{S}_{k}\bigr)\leq P
\end{equation}
}where $P$ is the maximum total power at the BS. Therefore, the
achievable rate optimization problem for the RIS-assisted MIMO BC
can be expressed as\begin{subequations}\label{eq:MIMO:BS:sumrate}{\small
\begin{align}
\underset{\{\mathbf{S}_{k}\succeq\mathbf{0}\},\boldsymbol{\theta}}{\maximize} & \quad\sum_{k=1}^{K}\log_{2}\frac{\Bigl|\mathbf{I}+\mathbf{H}_{\pi(k)}\bigl(\sum_{j\geq k}\mathbf{S}_{\pi(j)}\bigr)\mathbf{H}_{\pi(k)}\herm\Bigr|}{\Bigl|\mathbf{I}+\mathbf{H}_{\pi(k)}\bigl(\sum_{j>k}\mathbf{S}_{\pi(j)}\bigr)\mathbf{H}_{\pi(k)}\herm\Bigr|}\\
\st & \quad\sum\nolimits _{k=1}^{K}\tr\bigl(\mathbf{S}_{k}\bigr)\leq P\\
 & \quad\left|\theta_{l}\right|=1,\quad l=1,2,\ldots,N_{\mathrm{ris}}.
\end{align}
}\end{subequations} It is worth mentioning that the achievable sum-rate
in \eqref{eq:MIMO:BS:sumrate} is independent of the ordering of users
$\pi$ \cite{Vishwanath:duality_achievable:2003}. We remark that
the objective function of the above problem is neither convex nor
concave with the input covariance matrices and the phase shifts, and
thus directly solving \eqref{eq:MIMO:BS:sumrate} is difficult. In
\cite{Vishwanath:duality_achievable:2003}, Vishwanath \emph{et al.}
established what is now well-known as the \emph{BC-MAC duality}, and
showed that the achievable sum-rate of the MIMO \ac{BC} equals the
achievable rate of the dual Gaussian MIMO \ac{MAC}. As a result,
\eqref{eq:MIMO:BS:sumrate} is equivalent to\begin{subequations}\label{eq:MIMO:MAC:sumrate}{\small
\begin{align}
\underset{\{\bar{\mathbf{S}}_{k}\succeq\mathbf{0}\},\boldsymbol{\theta}}{\maximize} & \quad\log_{2}\Bigl|\mathbf{I}+\sum\nolimits _{k=1}^{K}\mathbf{H}_{k}\herm\bar{\mathbf{S}}_{k}\mathbf{H}_{k}\Bigr|\\
\st & \quad\sum\nolimits _{k=1}^{K}\tr\bigl(\bar{\mathbf{S}}_{k}\bigr)\leq P\\
 & \quad\left|\theta_{l}\right|=1,\quad l=1,2,\ldots,N_{\mathrm{ris}}
\end{align}
}\end{subequations} where $\mathbf{H}_{k}\herm$ is referred to
as the dual \ac{MAC} corresponding to $\mathbf{H}_{k}$ and $\bar{\mathbf{S}}_{k}\in\mathbb{C}^{n_{k}\times n_{k}}$
is the input covariance matrix of user $k$ in the dual MAC.\textcolor{blue}{}

\section{Alternating Optimization (AO)}

To solve \eqref{eq:MIMO:MAC:sumrate}, we propose an efficient \ac{AO}
method, which adjusts the covariance matrices and the RIS element
phase shifts in an alternating fashion. First, we propose an iterative
approach which optimizes all of the covariance matrices in the dual
MAC in a successive manner. Next, the optimal phase shift value for
each RIS element is obtained using a derived closed-form expression,
similar to \cite{zhang2019capacity}. We will analyze the computational
complexity of the proposed AO method in Subsection \ref{subsec:Computational-Complexity}.

\subsection{Covariance Matrix Optimization}

For a given $\boldsymbol{\theta}$, the achievable rate optimization
problem in \eqref{eq:MIMO:MAC:sumrate} can be simplified as\begin{subequations}\label{eq:MIMO:MAC:fixtheta}{\small
\begin{align}
\underset{\{\bar{\mathbf{S}}_{k}\succeq\mathbf{0}\}}{\maximize} & \quad\log_{2}\Bigl|\mathbf{I}+\sum\nolimits _{k=1}^{K}\mathbf{H}_{k}\herm\bar{\mathbf{S}}_{k}\mathbf{H}_{k}\Bigr|\\
\st & \quad\sum\nolimits _{k=1}^{K}\tr\bigl(\bar{\mathbf{S}}_{k}\bigr)\leq P.\label{eq:MAC:SPC}
\end{align}
}\end{subequations} This redefined optimization problem is convex
and thus it can be solved by off-the-shelf convex solvers. However,
we apply instead the dual decomposition method \cite{Yu:SumCapacity:MIMO_BC:Decomposition:2006}
to solve \eqref{eq:MIMO:MAC:fixtheta}, which is more efficient and
is described in Algorithm \ref{alg:DD:fixedtheta}.

The partial Lagrangian function of \eqref{eq:MIMO:MAC:fixtheta} is{\small
\begin{equation}
L(\mu,\{\bar{\mathbf{S}}_{k}\})=\ln\Bigl|\mathbf{I}+\sum_{k=1}^{K}\mathbf{H}_{k}\herm\bar{\mathbf{S}}_{k}\mathbf{H}_{k}\Bigr|-\mu\biggl[\sum_{k=1}^{K}\tr\bigl(\bar{\mathbf{S}}_{k}\bigr)-P\biggr]\label{eq:Lang_funct}
\end{equation}
}where $\mu$ is the Lagrangian multiplier for the constraint \eqref{eq:MAC:SPC}.
For mathematical convenience, we use the natural logarithm in the
previous expression without affecting the optimality of \eqref{eq:MIMO:MAC:fixtheta}.
For a given $\mu$, the dual objective is given as
\begin{equation}
g(\mu)=\max\;\bigl\{ L\bigl(\mu,\{\bar{\mathbf{S}}_{k}\}\bigr)\;\bigl|\;\{\bar{\mathbf{S}}_{k}\}\succeq\mathbf{0}\bigr\}\label{eq:dual}
\end{equation}
and its optimization can be performed by cyclically optimizing each
$\bar{S}_{k}$ in turn while keeping the other $\bar{S}_{j}$ $(j\ne k)$
fixed. To this end, let us consider the optimization of \eqref{eq:dual}
over $\bar{\mathbf{S}}_{k}$, which is expressed as
\begin{align}
\!\!\!\underset{\bar{\mathbf{S}}_{k}\succeq\mathbf{0}}{\maximize} & \quad\ln\Bigl|\mathbf{I}+\bar{\mathbf{H}}_{k}^{-1/2}\mathbf{H}_{k}\herm\bar{\mathbf{S}}_{k}\mathbf{H}_{k}\bar{\mathbf{H}}_{k}^{-1/2}\Bigr|-\mu\tr\bigl(\bar{\mathbf{S}}_{k}\bigr)\label{eq:Sbarmax}
\end{align}
where{\small 
\begin{equation}
\bar{\mathbf{H}}_{k}=\mathbf{I}+\sum\nolimits _{j=1,j\neq k}^{K}\mathbf{H}_{j}\herm\bar{\mathbf{S}}_{j}\mathbf{H}_{j}.
\end{equation}
}It is easy to see that the optimal solution to \eqref{eq:Sbarmax}
is given by{\small 
\begin{equation}
\bar{\mathbf{S}}_{k}^{\star}=\mathbf{V}_{k}\diag\Bigl(\Bigl[\bigl(\frac{1}{\mu}-\frac{1}{\sigma_{1}}\bigr)_{+},\bigl(\frac{1}{\mu}-\frac{1}{\sigma_{2}}\bigr)_{+},\ldots\bigl(\frac{1}{\mu}-\frac{1}{\sigma_{r}}\bigr)_{+}\Bigr]^{T}\Bigr)\mathbf{V}_{k}\herm\label{eq:Sk_opt}
\end{equation}
}where $\mathbf{H}_{k}\bar{\mathbf{H}}_{k}^{-1}\mathbf{H}_{k}\herm=\mathbf{V}_{k}\diag\Bigl(\sigma_{1},\sigma_{2},\ldots\sigma_{r}\Bigr)\mathbf{V}_{k}\herm$
is the \ac{EVD} of $\mathbf{H}_{k}\bar{\mathbf{H}}_{k}^{-1}\mathbf{H}_{k}\herm$
and $r=\rank(\mathbf{H}_{k})\le\min(N_{t},n_{k})$. Let $\{\bar{\mathbf{S}}_{k}^{\star}\}_{k=1}^{K}$
be the optimal solution of \eqref{eq:dual}. Next, the dual problem
is 
\begin{equation}
\minimize\ \{g(\mu)\;\bigl|\;\mu\geq0\}.\label{eq:dualprob}
\end{equation}
Since $P-\sum_{k=1}^{K}\tr\bigl(\bar{\mathbf{S}}_{k}^{\star}\bigr)$
is a subgradient of $g(\mu)$, the dual problem \eqref{eq:dualprob}
can be efficiently solved by a bisection search as outlined in Algorithm~\ref{alg:DD:fixedtheta}.
In particular, increase $\mu_{\min}$ if $P-\sum_{k=1}^{K}\tr\bigl(\bar{\mathbf{S}}_{k}^{\star}\bigr)<0$
and decrease $\mu_{\max}$ otherwise.

A possible upper limit of the bisection search for Algorithm~\ref{alg:DD:fixedtheta}
can be found as follows. From the \ac{KKT} condition of \eqref{eq:Sbarmax}
we have{\small
\[
\mathbf{H}_{k}\bigl(\bar{\mathbf{H}}_{k}^{-1/2}\bigr)\herm\bigl(\mathbf{I}+\bar{\mathbf{H}}_{k}^{-1/2}\mathbf{H}_{k}\herm\bar{\mathbf{S}}_{k}\mathbf{H}_{k}\bar{\mathbf{H}}_{k}^{-1/2}\bigr)^{-1}\bar{\mathbf{H}}_{k}^{-1/2}\mathbf{H}_{k}\herm+\mathbf{M}_{k}=\mu\mathbf{I}
\]
}where $\mathbf{M}_{k}\succeq\mathbf{0}$ is the Lagrangian multiplier
of the constraints $\bar{\mathbf{S}}_{k}\succeq\mathbf{0}$. Further,
this yields{\small
\[
\mathbf{H}_{k}\bigl(\bar{\mathbf{H}}_{k}^{-1/2}\bigr)\herm\bigl(\mathbf{I}+\bar{\mathbf{H}}_{k}^{-1/2}\mathbf{H}_{k}\herm\bar{\mathbf{S}}_{k}\mathbf{H}_{k}\bar{\mathbf{H}}_{k}^{-1/2}\bigr)^{-1}\bar{\mathbf{H}}_{k}^{-1/2}\mathbf{H}_{k}\herm\bar{\mathbf{S}}_{k}=\mu\bar{\mathbf{S}}_{k}
\]
}and thus{\small
\begin{gather}
\tr\bigl(\bigl(\mathbf{I}+\bar{\mathbf{H}}_{k}^{-1/2}\mathbf{H}_{k}\herm\bar{\mathbf{S}}_{k}\mathbf{H}_{k}\bar{\mathbf{H}}_{k}^{-1/2}\bigr)^{-1}\bar{\mathbf{H}}_{k}^{-1/2}\mathbf{H}_{k}\herm\nonumber \\
\times\bar{\mathbf{S}}_{k}\mathbf{H}_{k}\bigl(\bar{\mathbf{H}}_{k}^{-1/2}\bigr)\herm\bigr)=\mu\tr\bigl(\bar{\mathbf{S}}_{k}\bigr).
\end{gather}
}Note that $\tr\bigl(\bigl(\mathbf{I}+\mathbf{A}\bigr)^{-1}\mathbf{A}\bigr)=\tr\bigl(\bigl(\mathbf{I}+\mathbf{A}^{-1}\bigr)^{-1}\bigr)\leq N_{t}$
and thus the above equality implies $\mu\tr\bigl(\bar{\mathbf{S}}_{k}\bigr)\leq N_{t}$.
Combining this inequality for all users, we have $\mu\leq KN_{t}/P$.
Hence, setting $\mu_{max}=KN_{t}/P$ in Algorithm \ref{alg:DD:fixedtheta}
guarantees finding the optimal solution to \eqref{eq:MIMO:MAC:fixtheta}.

\subsection{RIS Optimization}

For fixed $\{\bar{\mathbf{S}}_{k}\}_{k=1}^{K}$ and $\{\theta_{m},m\neq l\}_{m=1}^{N_{\mathrm{ris}}}$,
the optimization problem in \eqref{eq:MIMO:MAC:sumrate} with respect
to $\theta_{l}$ can be explicitly written as\begin{subequations}\label{eq:MIMO:MAC:fixcovar}{\small
\begin{align}
\underset{\theta_{l}}{\maximize} & \quad\log_{2}\Bigl|\mathbf{I}+\sum\nolimits _{k=1}^{K}\mathbf{H}_{k}\herm\bar{\mathbf{S}}_{k}\mathbf{H}_{k}\Bigl|\\
\st & \quad\left|\theta_{l}\right|=1.\label{eq:phaseshift}
\end{align}
}\end{subequations}To proceed further, we present the objective
of \eqref{eq:MIMO:MAC:fixcovar} as $\log_{2}\Bigl|\mathbf{A}_{l}+\theta_{l}\mathbf{B}_{l}+\theta_{l}^{\ast}\mathbf{B}_{l}\herm\Bigr|$,
where{\small
\begin{gather}
\mathbf{A}_{l}=\mathbf{I}+\sum\nolimits _{k=1}^{K}\bigl(\mathbf{D}_{k}\herm+\sum\nolimits _{\underset{m\neq l}{m=1}}^{N_{\mathrm{ris}}}\theta_{m}^{*}\mathbf{u}_{m}\herm\mathbf{g}_{k,m}\herm)\bar{\mathbf{S}}_{k}\nonumber \\
\times\bigl(\mathbf{D}_{k}+\sum\nolimits _{\underset{n\neq l}{n=1}}^{N_{\mathrm{ris}}}\theta_{n}\mathbf{g}_{k,n}\mathbf{u}_{n})+\sum\nolimits _{k=1}^{K}\mathbf{u}_{l}\herm\mathbf{g}_{k,l}\herm\bar{\mathbf{S}}_{k}\mathbf{g}_{k,l}\mathbf{u}_{l},\label{eq:Equ_Al}
\end{gather}
\begin{equation}
\mathbf{B}_{l}=\Bigl[\sum\nolimits _{k=1}^{K}\bigl(\mathbf{D}_{k}\herm+\sum\nolimits _{\underset{m\neq l}{m=1}}^{N_{\mathrm{ris}}}\theta_{m}^{*}\mathbf{u}_{m}\herm\mathbf{g}_{k,m}\herm\bigr)\bar{\mathbf{S}}_{k}\mathbf{g}_{k,l}\Bigr]\mathbf{u}_{l},\label{eq:Equ_Bl}
\end{equation}
}$\mathbf{U}=[\mathbf{u}_{1}^{T}\;\mathbf{u}_{2}^{T}\;\cdots\;\mathbf{u}_{N_{\mathrm{ris}}}^{T}]^{T}$
and $\mathbf{G}_{k}=[\mathbf{g}_{k,1}\;\mathbf{g}_{k,2}\;\cdots\;\mathbf{g}_{k,N_{\mathrm{ris}}}]$.

The optimal solution to \eqref{eq:MIMO:MAC:fixcovar} is then given
by \cite{zhang2019capacity}
\begin{algorithm}[t]
{\small\caption{Dual decomposition for solving \eqref{eq:MIMO:MAC:fixtheta}.\label{alg:DD:fixedtheta}}

\SetAlgoNoLine
\DontPrintSemicolon
\LinesNumbered

\KwIn{ $\mu_{\min}=0$, $\mu_{\max}>0$, $\epsilon>0$: desired
accuracy. }

\Repeat{$\mu_{\max}-\mu_{\min}<\epsilon$ }{

Set $\mu=\frac{\mu_{\max}+\mu_{\min}}{2}$ and $k=0$

\Repeat{convergence of \eqref{eq:Lang_funct} }{

Set $k\leftarrow(k\!\!\mod K)+1$

Compute $\bar{\mathbf{S}}_{k}^{\star}$ according to \eqref{eq:Sk_opt}

}

\lIf{$P<\sum_{k=1}^{K}\tr\bigl(\bar{\mathbf{S}}_{k}^{\star}\bigr)$
}{ Set $\mu_{\min}=\mu$ }\lElse{ Set $\mu_{\max}=\mu$}

}

}
\end{algorithm}
\begin{equation}
\theta_{l}^{\star}=\exp(-j\arg(\sigma_{l})),\label{eq:opttheta}
\end{equation}
where $\sigma_{l}$ is the only non-zero eigenvalue of $\mathbf{A}_{l}^{-1}\mathbf{B}_{l}$
(it can be observed from \eqref{eq:Equ_Bl} that the rank of $\mathbf{B}_{l}$
is equal to 1).

\subsection{Overall AO Method}

The overall AO algorithm description is given in Algorithm~\ref{alg:AO}.
At first, we compute the optimal covariance matrices for all users,
$\{\bar{\mathbf{S}}_{k}\}_{k=1}^{K}$. Next, we sequentially obtain
the optimal phase shift value for each RIS element. These two optimization
steps constitute one \emph{outer iteration} of Algorithm~\ref{alg:AO}.

It is obvious that each iteration of the AO algorithm increases the
achievable sum-rate. Also, the solution in each iteration of the AO
method is unique and the feasible set is compact. Thus, the convergence
of the AO method to a stationary solution is guaranteed. However,
since the problem \eqref{eq:MIMO:BS:sumrate} is non-convex, we cannot
claim that the obtained solution is globally optimal.

\subsection{Computational Complexity\label{subsec:Computational-Complexity}}

In this subsection, the computational complexity is obtained by counting
the required number of complex multiplications. The complexity of
the AO is determined by the computation of the covariance matrices
$\{\bar{\mathbf{S}}_{k}\}_{k=1}^{K}$ and the RIS phase shifts $\{\theta_{m}\}_{m=1}^{N_{\mathrm{ris}}}$
in Algorithm \ref{alg:AO}. In the following complexity derivation,
for ease of exposition we assume that all the users have the same
number of antennas, i.e., $n_{k}=N_{r}$ for all $k=1,2,\dots,K$.
At first, we need to compute all the users' channel matrices. To compute
$\mathbf{F}(\boldsymbol{\theta})\mathbf{U}$ requires $N_{\mathrm{ris}}N_{t}$
multiplications and it is common for all users. To form $\mathbf{G}_{k}\mathbf{F}(\boldsymbol{\theta})\mathbf{U}$,
we require $N_{\mathrm{ris}}N_{t}N_{r}$ further multiplications per
user, so the complexity of calculating all of the user's channel matrices
is $\mathcal{O}(KN_{\mathrm{ris}}N_{t}N_{r})$. To reduce the complexity
of computing $\bar{\mathbf{H}}_{k}$, instead of following (12) we
compute and store $\mathbf{H}_{\mathrm{sum}}=\mathbf{I}+\sum_{j=1}^{K}\mathbf{H}_{j}\herm\bar{\mathbf{S}}_{j}\mathbf{H}_{j}$,
which requires $\mathcal{O}(KN_{t}N_{r}^{2}+KN_{t}^{2}N_{r})$ multiplications.
Then $\bar{\mathbf{H}}_{k}^{-1}=(\mathbf{H}_{\mathrm{sum}}-\mathbf{H}_{k}\herm\bar{\mathbf{S}}_{k}\mathbf{H}_{k})^{-1}$
has a complexity of $\mathcal{O}(N_{t}^{3})$ and $\mathbf{H}_{k}\bar{\mathbf{H}}_{k}^{-1}\mathbf{H}_{k}\herm$
has a complexity of $\mathcal{O}(N_{t}N_{r}^{2}+N_{t}^{2}N_{r})$.
The \ac{EVD} of $\mathbf{H}_{k}\bar{\mathbf{H}}_{k}^{-1}\mathbf{H}_{k}\herm$
requires $\mathcal{O}(N_{r}^{3})$ multiplications, while the complexity
of computing $\bar{\mathbf{S}}_{k}^{\star}$ is $\mathcal{O}(N_{r}^{3})$.
To recalculate $\mathbf{H}_{\mathrm{sum}}=\bar{\mathbf{H}}_{k}+\mathbf{H}_{k}\herm\bar{\mathbf{S}}_{k}\mathbf{H}_{k}$,
we need $\mathcal{O}(N_{t}N_{r}^{2}+N_{t}^{2}N_{r})$ multiplications.
Finally, the complexity of calculating $\{\bar{\mathbf{S}}_{k}\}_{k=1}^{K}$
may be expressed as $\mathcal{O}(KN_{\mathrm{ris}}N_{t}N_{r}+KN_{t}N_{r}^{2}+KN_{t}^{2}N_{r}+LI(N_{t}^{3}+2N_{t}N_{r}^{2}+2N_{t}^{2}N_{r}+2N_{r}^{3}))=\mathcal{O}(KN_{\mathrm{ris}}N_{t}N_{r}+LI(N_{t}^{3}+N_{t}N_{r}^{2}+N_{t}^{2}N_{r}+N_{r}^{3}))$,
where $L$ is the required number of outer iterations (i.e., lines
1-9) in Algorithm \ref{alg:DD:fixedtheta}, and $I$ is the average
number of iterations required for the optimization of the covariance
matrices (i.e., rows 3 to 6) in Algorithm \ref{alg:DD:fixedtheta}.
In our case, $L$ is the smallest integer that satisfies $\mu_{\max}/2^{L}<\epsilon$.
From numerical experiments, we have observed that $I<2K$ is usually
sufficient to attain a difference between two consecutive values of
\eqref{eq:Lang_funct} that is lower than $10^{-6}$. 
\begin{algorithm}[t]
{\small\caption{AO algorithm for solving \eqref{eq:MIMO:MAC:sumrate}. \label{alg:AO}}

\SetAlgoNoLine
\DontPrintSemicolon
\LinesNumbered 

\KwIn{ $|\theta_{l}|=1,l=1,2,\ldots,N_{\mathrm{ris}}$ }

\Repeat{convergence }{

Solve for $\{\bar{\mathbf{S}}_{k}^{\star}\}_{k=1}^{K}$ using \textbf{Algorithm
\ref{alg:DD:fixedtheta}} \\

\For{$l=1,2,\ldots,N_{\mathrm{ris}}$}{

Solve for $\theta_{l}^{\star}$ using \eqref{eq:opttheta}

}

}

}
\end{algorithm}

The complexity of computing the optimal RIS phase shifts is primarily
dependent on \eqref{eq:Equ_Al} and \eqref{eq:Equ_Bl}. Let us define
$\mathbf{C}_{k}=\mathbf{H}_{k}-\theta_{l}\mathbf{g}_{k,l}\mathbf{u}_{l}$
to simplify the complexity derivation. It is easy to see that we need
$\mathcal{O}(N_{t}N_{r})$ multiplications to obtain $\mathbf{C}_{k}$
from $\mathbf{H}_{k}$. The complexity of computing the matrix product
$\mathbf{C}_{k}\bar{\mathbf{S}}_{k}\mathbf{C}_{k}$ is $\mathcal{O}(N_{t}N_{r}^{2}+N_{t}^{2}N_{r})$.
In a similar manner, the complexity of $\mathbf{u}_{l}\herm\mathbf{g}_{k,l}\herm\bar{\mathbf{S}}_{k}\mathbf{g}_{k,l}\mathbf{u}_{l}$
is equal to $\mathcal{O}(N_{t}N_{r}^{2}+N_{t}^{2}N_{r})$. Hence,
the complexity of computing $\mathbf{A}_{l}$ in \eqref{eq:Equ_Al}
is $\mathcal{O}(KN_{t}N_{r}^{2}+KN_{t}^{2}N_{r})$. Also, we need
$\mathcal{O}(KN_{t}^{2}N_{r})$ more multiplications to obtain $\mathbf{B}_{l}$
in \eqref{eq:Equ_Bl}. Inverting $\mathbf{A}_{l}$ requires $\mathcal{O}(N_{t}^{3})$
multiplications. The same complexity is required for computing $\mathbf{A}_{l}^{-1}\mathbf{B}_{l}$
and for obtaining the \ac{EVD} of that product. The complexity of
computing a single RIS phase shift is $\mathcal{O}(KN_{t}N_{r}^{2}+KN_{t}^{2}N_{r}+N_{t}^{3})$,
which gives a total~of $\mathcal{O}(KN_{\mathrm{ris}}N_{t}N_{r}^{2}+KN_{\mathrm{ris}}N_{t}^{2}N_{r}+N_{\mathrm{ris}}N_{t}^{3})$
for the whole RIS.

In summary, the computational complexity of one outer iteration (i.e.,
lines 1 to 6 in Algorithm \ref{alg:AO}) of the AO algorithm is given
by{\small
\begin{align}
C_{\mathrm{AO}}= & \,\mathcal{O}(KN_{\mathrm{ris}}N_{t}N_{r}^{2}+KN_{\mathrm{ris}}N_{t}^{2}N_{r}+N_{\mathrm{ris}}N_{t}^{3}\nonumber \\
 & \qquad+LI(N_{t}^{3}+N_{t}N_{r}^{2}+N_{t}^{2}N_{r}+N_{r}^{3})).\label{eq:AO_compl}
\end{align}
}\vspace{-0.5cm}

\section{Simulation Results}

In this section, we evaluate the achievable rate of the proposed AO
algorithm with the aid of Monte Carlo simulations. The study is conducted
for a typical multi-user propagation environment in three different
scenarios: (i) where only the direct link (i.e., the first term in
\eqref{eq:Hk_equ}) is present; (ii) where only the link via the RIS
(i.e., the second term in \eqref{eq:Hk_equ}) is present; and (iii)
where both of these links are present. In order to better quantify
the gains of the proposed AO method, we present the achievable sum-rate
results for different numbers of users and for different numbers of
transmit antennas.

The positions of the BS, the RIS and the users are specified by a
three-dimensional (3D) Cartesian coordinate system. The BS ULA is
placed parallel to the \emph{y}-axis and the position of its midpoint
is set as $(0,l_{t},h_{t})$. The RIS is located in the \emph{xz}-plane
and the position of its midpoint is $(d_{\mathrm{ris}},0,h_{\mathrm{ris}})$.
For simplicity, we assume that all of the users' ULAs are parallel
to the \mbox{\emph{y}-axis} and the midpoint of the \emph{k}-th
user's ULA is located at $(d_{k},l_{k},h_{k})$. For the considered
system geometry, the distance between the midpoint of the BS ULA and
the midpoint of the RIS is $d_{t,\mathrm{ris}}=\sqrt{d_{\mathrm{ris}}^{2}+l_{t}^{2}+(h_{t}-h_{\mathrm{ris}})^{2}}$,
the distance between the midpoint of the RIS and the midpoint of the
\mbox{\emph{k}-th} user's ULA is $d_{\mathrm{ris},k}=\sqrt{(d_{\mathrm{ris}}-d_{k})^{2}+l_{k}^{2}+(h_{\mathrm{ris}}-h_{k})^{2}}$,
and the distance between the midpoint of the BS ULA and the midpoint
of the \emph{k}-th user's ULA is $d_{t,k}=\sqrt{d_{k}^{2}+(l_{t}-l_{k})^{2}+(h_{t}-h_{k})^{2}}$.

\textcolor{black}{In the following simulations, all of the channel
matrices are modeled according to the Rician fading channel model
with Rician factor equal to 1, as specified in \cite{perovic2020achievable}.
Also, we neglect spatial correlation among the elements of matrices
$\mathbf{U}$ and $\mathbf{G}_{k}$. The distance-dependent path loss
}for the direct link of the \mbox{\emph{k}-th} user is $\beta_{\mathrm{DIR},k}=(4\pi/\lambda)^{2}d_{t,k}^{\alpha_{\mathrm{DIR}}}$,
where $\alpha_{\mathrm{DIR}}$ denotes the path loss exponent of the
direct link. The far-field \ac{FSPL} for the RIS link of the \emph{k}-th
user $\beta_{\mathrm{RIS},k}$\textcolor{black}{{} is equal to $\beta_{\mathrm{RIS},k}^{-1}=G_{t}G_{r}\lambda^{4}\cos\gamma_{t}\cos\gamma_{r}/(256\pi^{2}d_{t,\mathrm{ris}}^{2}d_{\mathrm{ris},k}^{2})$,
where $\gamma_{t}$ is the angle between the incident wave propagation
direction and the normal to the RIS, and $\gamma_{r}$ is the angle
between the normal to the RIS and the reflected wave propagation direction
\cite[Eq. (7), (9)]{tang2020wireless}. Hence, we have $\cos\gamma_{t}=l_{t}/d_{t,\mathrm{ris}}$
and $\cos\gamma_{r}=l_{k}/d_{\mathrm{ris},k}$. Here $G_{t}$ and
$G_{r}$ represent the transmit and receive antenna gains respectively;
these values are both set to 2, since we assume that these antennas
radiate/sense signals to/from the relevant half space \cite{tang2020wireless}.
In this paper, $\sqrt{\beta_{\mathrm{DIR},k}^{-1}/N_{0}}$ and $\sqrt{\beta_{\mathrm{RIS},k}^{-1}/N_{0}}$
are embedded as scaling factors in $\mathbf{D}_{k}$ and $\mathbf{G}_{k}$,
respectively.}

In the following simulation setup, the parameters are $f=2\,\mathrm{GHz}$
(i.e., $\lambda=15\,\mathrm{cm}$), $s_{t}=s_{r}=s_{\mathrm{ris}}=\lambda/2=7.5\,\mathrm{cm}$,
$l_{t}=20\,\mathrm{m}$, $h_{t}=10\,\mathrm{m}$, $d_{\mathrm{ris}}=30\,\mathrm{m}$,
$h_{\mathrm{ris}}=5\,\mathrm{m}$, $N_{t}=8$, $\alpha_{\mathrm{DIR}}=3$,
$P=1\,\mathrm{W}$, and $N_{0}=-110\thinspace\mathrm{dB}$. The RIS
consists of $N_{\mathrm{ris}}=225$ elements placed in a \mbox{$15\times15$}
square formation. As in the previous section, we assume that all users
are equipped with $N_{r}=2$ antennas. The users' coordinates are
randomly selected such that $d_{k}$ is chosen from a uniform distribution
between $200\,\mathrm{m}$ to $500\,\mathrm{m}$ with a resolution
of $2\,\mathrm{m}$, $l_{k}$ is chosen from a uniform distribution
between $0$ to $70\,\mathrm{m}$ with a resolution of $1\,\mathrm{m}$,
and $h_{k}$ is chosen from a uniform distribution between $1.5\,\mathrm{m}$
to $2\,\mathrm{m}$ with a resolution of $1\,\mathrm{cm}$.\textcolor{blue}{{}
}\textcolor{black}{All results are averaged over 1000 independent
channel realizations.}
\begin{figure}[t]
\centering{}\input{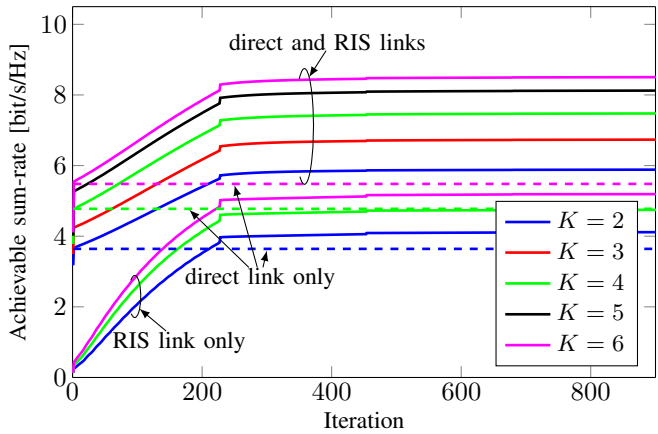}\textcolor{black}{\caption{Achievable sum-rate for the proposed AO method with direct and RIS
links, direct link only, and RIS link only. \label{fig:Ach-rate-AO}}
}
\end{figure}

The achievable sum-rate for the proposed AO method for the three cases
where the channel consists of the direct link only, the RIS link only,
and both the direct and RIS links, are shown in Fig. \ref{fig:Ach-rate-AO}.
As expected, the achievable sum-rate in multi-user communications
is higher when the direct link is present, similar to observations
regarding the achievable rate in point-to-point communications which
were reported in \cite{perovic2020achievable}. Moreover, the achievable
sum-rate increases with the number of users $K$, and this increase
is more substantial when the direct link is present. It seems that
the lack of amplitude adjustment capability prevents the RIS from
achieving a significant suppression of the multi-user interference.
Therefore, higher achievable sum-rate gains can be expected when a
part of a signal is transmitted via the direct link, since the BS
with its amplitude adjustment capabilities is better equipped to suppress
the aforementioned interference. In addition, the achievable sum-rate
increase starts to noticeably decline\textcolor{blue}{{} }when the number
of users is greater than four. The reason for this is that the channel
matrix of the aggregate BC between the BS and all users has the maximum
rank when the number of users $K$ is greater than or equal to four.
In other words, the rank of the channel matrix (i.e., number of degrees
of freedom) does not change if $K\ge4$ and this leads to \textcolor{black}{a
reduced }gain in the achievable sum-rate when increasing $K$ beyond
4.
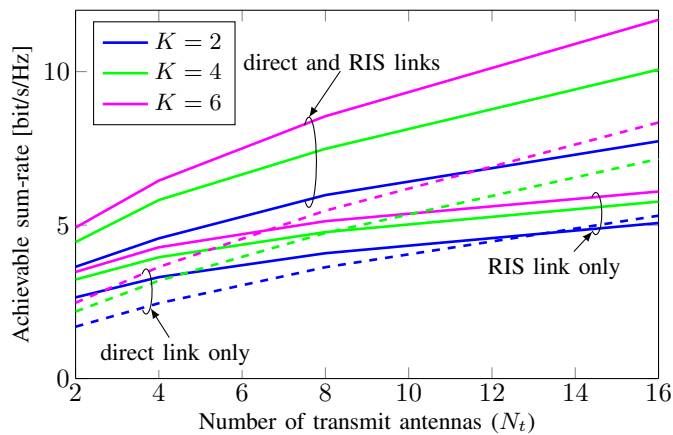
\begin{figure}[t]
\centering{}\input{figures/sweep_AO_new_V2.tex}\caption{Achievable sum-rate versus the number of transmit antennas ($N_{t}$).\label{fig:AO_DIR}}
\end{figure}

In Fig. \ref{fig:AO_DIR}, we show the achievable sum-rate for the
proposed AO method versus the number of transmit antennas $N_{t}$.
The achievable sum-rate curves have an approximately logarithmic shape.
Also, it can be observed that the achievable sum-rate increases with
the number of users. However, it seems that this increase gradually
declines with the increase of the number of users. At the same time,
the achievable sum-rate increases with the number of transmit antennas.
For example, for 6 users and 2 transmit antennas, a 99\,\% increase
in the achievable sum-rate is obtained by adding the RIS to the multi-user
system.\textcolor{black}{}

\section{Conclusion}

In this paper, we proposed an AO algorithm for the achievable sum-rate
optimization in a multi-user \ac{BC} that is equipped with an \ac{RIS}.
The algorithm is based on the well-known \ac{BC}-\ac{MAC} duality
for multi-user systems. The users\textquoteright{} covariance matrices
were optimized by a dual decomposition method, while the optimal RIS
phase shifts were computed by using a derived closed-form expression.
Also, we presented a computation complexity analysis for the proposed
AO algorithm. Simulation results show that adding the RIS can significantly
improve the achievable sum-rate in a BC.

\bibliographystyle{IEEEtran}
\bibliography{IEEEabrv,IEEEexample,references}

\end{document}

%% file: acro.tex
\acrodef{AO}{alternating optimization}
\acrodef{ASP}{antenna separation product}
\acrodef{AWGN}{additive white Gaussian noise}
\acrodef{BC}{broadcast channel}
\acrodef{BEP}{bit error probability}
\acrodef{BER}{bit error rate}
\acrodef{BF-MIMO}[BF\mbox{-}MIMO]{beamforming MIMO}
\acrodef{BF}{beamforming}
\acrodef{BS}{base station}
\acrodef{bpcu}{bits per channel use}
\acrodef{CP}{cyclic prefix}
\acrodef{CSI}{channel state information}
\acrodef{CSIR}{channel state information at RX}
\acrodef{SSK}{space shift keying}
\acrodef{CSIT}{channel state information at TX}
\acrodef{DCMC}{discrete\mbox{-}input continuous\mbox{-}output memoryless channel}
\acrodef{DFT}{discrete Fourier transform}
\acrodef{DL-TR-GSM}{dual-layered transmit-receive \acl{GSM}}
\acrodef{DLT}{dual-layered transmission}
\acrodef{DPC}{dirty paper coding}
\acrodef{EGC}{equal gain combining}
\acrodef{EM}{electromagnetic}
\acrodef {EVD}{eigenvalue decomposition}
\acrodef{FSPL}{free space path loss}
\acrodef{FFT}{fast Fourier transform}
\acrodef{FDE}{frequency domain equalization}
\acrodef{GRSM}{generalized \acl{RSM}}
\acrodef{GSM}{generalized \acl{SM}}
\acrodef{IFFT}{invserse fast Fourier transform}
\acrodef{ICI}{inter-channel interference}
\acrodef{iid}[i.i.d.]{independent and identically distributed}
\acrodef{IQ}{in\mbox{-}phase and quadrature}
\acrodef{ISI}{intersymbol interference}
\acrodef{ISI-free}[ISI\mbox{-}free]{intersymbol interference free}
\acrodef{LIS}{large intelligent surface}
\acrodef{LOS}{line\mbox{-}of\mbox{-}sight}
\acrodef{KKT}{Karush\mbox{-}Kuhn\mbox{-}Tucker} 
\acrodef{MAC}{multiple-access channel}
\acrodef{mmWave}{millimeter-wave}
\acrodef{MIMO}{multiple\mbox{-}input multiple\mbox{-}output}
\acrodef{MISO}{multiple\mbox{-}input single\mbox{-}output}
\acrodef{ML}{maximum likelihood}
\acrodef{MRC}{maximal ratio combining}
\acrodef{MMSE}{minimum mean square error}
\acrodef{MU-TR-GSM}{multiuser transmit-receive  \acl{GSM} }
\acrodef{NCSIT}{no channel state information at TX}
\acrodef{NLOS}{non\mbox{-}\acs{LOS}} 
\acrodef{NOMA}{non-orthogonal multiple access}
\acrodef{OFDM}{orthogonal frequency division multiplexing}
\acrodef{OFDMA}{orthogonal frequency division multiple access}
\acrodef{PA}{power amplifier}
\acrodef{PAE}{power added efficiency}
\acrodef{PAPR}{peak\mbox{-}to\mbox{-}average power ratio}
\acrodef{PDF}{probability density function}
\acrodef{PEP}{pairwise error probability}
\acrodefplural{PEP}{pairwise error probabilities}
\acrodef{PGM}{projected gradient method}
\acrodef{APGM}{accelerated projected gradient method}
\acrodef{PMP}{probability mass function}
\acrodef{PSM}{precoding-aided spatial modulation}
\acrodef{QSM}{quadrature spatial modulation}
\acrodef{RC}{reorganization computation}
\acrodef{RIS}{reconfigurable intelligent surface}
\acrodef{RSM}{receive spatial modulation}
\acrodef{RX}{receiver}
\acrodef{SEP}{symbol error probability}
\acrodef{SER}{symbol error rate}
\acrodef{SINR}{signal-to-interference-plus-noise ratio}
\acrodef{SISO}{single-input single-output}
\acrodef{SM}{spatial modulation}
\acrodef{SMX-MIMO}[SMX\mbox{-}MIMO]{spatial multiplexing MIMO}
\acrodef{SMX}{spatial multiplexing}
\acrodef{SNR}{signal-to-noise ratio}
\acrodef{SC}{single carrier}
\acrodef{SVD}{singular value decomposition}
\acrodef{SPST}{single pole single-throw}
\acrodef{SU}{secondary user}
\acrodef{TDE}{time domain equalization}
\acrodef{TX}{transmitter}
\acrodef{ULA}{uniform linear array}
\acrodef{URA}{uniform rectangular array}
\acrodef{VGA}{variable gain amplifier}
\acrodef{ZF}{zero-forcing}
\acrodef{ZMCG}{zero-mean complex Gaussian}

%% file: figures/sweep_AO_new_V2.tex
%
\pgfplotsset{compat=newest}
\usetikzlibrary{plotmarks}
\usepgfplotslibrary{patchplots}

\setlength\figurewidth{7.75cm}
\setlength\figureheight{5.35cm}
%
\begin{tikzpicture}

\begin{axis}[%
width=\figurewidth,
height=0.916\figureheight,
at={(0\figurewidth,0\figureheight)},
scale only axis,
separate axis lines,
every outer x axis line/.append style={black},
every x tick label/.append style={font=\color{black}},
xmin=2,
xmax=16,
xlabel={Number of transmit antennas ($N_t$)},
every outer y axis line/.append style={black},
every y tick label/.append style={font=\color{black}},
ymin=0,
ymax=12,
ylabel={Achievable sum-rate [bit/s/Hz]},
axis background/.style={fill=white},
legend style={at={(0.03,0.97)},anchor=north west,legend cell align=left,align=left,draw=black}
]
\addplot [color=blue,solid,line width=1.0pt]
  table[row sep=crcr]{%
2	3.64106139503503\\
4	4.56767701193939\\
8	5.97799614458297\\
16	7.73241428015873\\
};
\addlegendentry{$K=2$};

\addplot [color=green,solid,line width=1.0pt]
  table[row sep=crcr]{%
2	4.44599592025648\\
4	5.8165148693634\\
8	7.48517914345564\\
16	10.0664852932966\\
};
\addlegendentry{$K=4$};

\addplot [color=mycolor1,solid,line width=1.0pt]
  table[row sep=crcr]{%
2	4.91926304523992\\
4	6.44928332904681\\
8	8.54758655052263\\
16	11.6887317881267\\
};
\addlegendentry{$K=6$};

\addplot [color=blue,solid,line width=1.0pt]
  table[row sep=crcr]{%
2	2.64487647458151\\
4	3.30376340291969\\
8	4.0838234453572\\
16	5.06763484220877\\
};

\addplot [color=green,solid,line width=1.0pt]
  table[row sep=crcr]{%
2	3.22578326797425\\
4	3.95245123573548\\
8	4.7729585707792\\
16	5.76501257530757\\
};

\addplot [color=mycolor1,solid,line width=1.0pt]
  table[row sep=crcr]{%
2	3.47378945837927\\
4	4.27706947487353\\
8	5.12825699856713\\
16	6.09395834044016\\
};

\addplot [color=blue,dashed,line width=1.0pt]
  table[row sep=crcr]{%
2	1.68999130739372\\
4	2.45451230780658\\
8	3.62781114561236\\
16	5.30659611175384\\
};

\addplot [color=green,dashed,line width=1.0pt]
  table[row sep=crcr]{%
2	2.18445095761167\\
4	3.18845354474738\\
8	4.74342311395666\\
16	7.15604960746516\\
};

\addplot [color=mycolor1,dashed,line width=1.0pt]
  table[row sep=crcr]{%
2	2.46995209841521\\
4	3.63492396933503\\
8	5.47053182759515\\
16	8.34253949865414\\
};

\node[align=center,fill=white,inner sep=2pt] (D1) at (axis cs: 13.6,3.6) { RIS  link only };
\draw (axis cs: 14.4, 5.9) arc (130:-125: 0.08cm and  0.28cm);
\draw[->,>=latex] (D1) -- (axis cs: 14.42,4.8);

\draw (axis cs: 7.5, 8.25) arc (125:-120: 0.1cm and  0.6cm);
\node[align=center,fill=white,inner sep =0pt] (D2) at (axis cs: 8.5,10.3) { direct and RIS links };
\draw[->,>=latex] (D2) -- (axis cs: 7.6,8.5);

\node[align=center,fill=white,inner sep=2pt] (D3) at (axis cs: 4.5,0.8) { direct link only };
\draw (axis cs: 3.6, 3.4) arc (130:-125: 0.08cm and  0.3cm);
\draw[->,>=latex] (D3) -- (axis cs: 3.8,2.3);

\end{axis}
\end{tikzpicture}%

%% file: Multiuser_AO_Rev_v1.bbl
\begin{thebibliography}{10}
\providecommand{\url}[1]{#1}
\csname url@samestyle\endcsname
\providecommand{\newblock}{\relax}
\providecommand{\bibinfo}[2]{#2}
\providecommand{\BIBentrySTDinterwordspacing}{\spaceskip=0pt\relax}
\providecommand{\BIBentryALTinterwordstretchfactor}{4}
\providecommand{\BIBentryALTinterwordspacing}{\spaceskip=\fontdimen2\font plus
\BIBentryALTinterwordstretchfactor\fontdimen3\font minus
  \fontdimen4\font\relax}
\providecommand{\BIBforeignlanguage}[2]{{%
\expandafter\ifx\csname l@#1\endcsname\relax
\typeout{** WARNING: IEEEtran.bst: No hyphenation pattern has been}%
\typeout{** loaded for the language `#1'. Using the pattern for}%
\typeout{** the default language instead.}%
\else
\language=\csname l@#1\endcsname
\fi
#2}}
\providecommand{\BIBdecl}{\relax}
\BIBdecl

\bibitem{di2019smart}
M.~Di~Renzo \emph{et~al.}, ``Smart radio environments empowered by
  reconfigurable {AI} meta-surfaces: An idea whose time has come,''
  \emph{EURASIP J. Wireless Commun. and Netw.}, vol. 2019, no.~1, pp. 1--20,
  2019.

\bibitem{di2020smart}
------, ``Smart radio environments empowered by reconfigurable intelligent
  surfaces: How it works, state of research, and road ahead,'' \emph{{IEEE} J.
  Sel. Areas Commun.}, vol.~38, no.~11, pp. 2450--2525, Nov. 2020.

\bibitem{perovic2020achievable}
N.~S. Perovi{\'c} \emph{et~al.}, ``Achievable rate optimization for {MIMO}
  systems with reconfigurable intelligent surfaces,'' \emph{{IEEE} Trans.
  Wireless Commun.}, vol.~20, no.~6, pp. 3865--3882, Jun. 2021.

\bibitem{zhang2019capacity}
S.~Zhang and R.~Zhang, ``{Capacity characterization for intelligent reflecting
  surface aided MIMO communication},'' \emph{{IEEE} J. Sel. Areas Commun.},
  vol.~38, no.~8, pp. 1823--1838, Aug. 2020.

\bibitem{perovic2019channel}
N.~S. Perovi{\'c} \emph{et~al.}, ``Channel capacity optimization using
  reconfigurable intelligent surfaces in indoor {mmWave} environments,'' in
  \emph{Proc. IEEE Int. Conf. on Communications (ICC)}, 2020, pp. 1--7.

\bibitem{perovic2020optimization}
------, ``Optimization of {RIS}-aided {MIMO} systems via the cutoff rate,''
  \emph{{IEEE} Wireless Commun. Lett.}, 2021, {Early access}.

\bibitem{zhang2020intelligent}
S.~Zhang and R.~Zhang, ``Intelligent reflecting surface aided multi-user
  communication: Capacity region and deployment strategy,'' \emph{{IEEE} Trans.
  Wireless Commun.}, 2021, {Early access}.

\bibitem{guo2019weighted}
H.~Guo \emph{et~al.}, ``Weighted sum-rate maximization for intelligent
  reflecting surface enhanced wireless networks,'' in \emph{Proc. IEEE Global
  Communications Conference (GLOBECOM)}, 2019, pp. 1--6.

\bibitem{kammoun2020asymptotic}
Q.-U.-A. Nadeem \emph{et~al.}, ``Asymptotic max-min {SINR} analysis of
  reconfigurable intelligent surface assisted {MISO} systems,'' \emph{{IEEE}
  Trans. Wireless Commun.}, vol.~19, no.~12, pp. 7748--7764, Dec. 2020.

\bibitem{pan2020multicell}
C.~Pan \emph{et~al.}, ``Multicell {MIMO} communications relying on intelligent
  reflecting surfaces,'' \emph{{IEEE} Trans. Wireless Commun.}, vol.~19, no.~8,
  pp. 5218--5233, Aug. 2020.

\bibitem{pan2020intelligent}
------, ``Intelligent reflecting surface aided {MIMO} broadcasting for
  simultaneous wireless information and power transfer,'' \emph{{IEEE} J. Sel.
  Areas Commun.}, vol.~38, no.~8, pp. 1719--1734, Aug. 2020.

\bibitem{Weingarten:CapacityRegion:MU_MIMO:2006}
H.~Weingarten \emph{et~al.}, ``The capacity region of the {Gaussian}
  multiple-input multiple-output broadcast channel,'' \emph{{IEEE} Trans. Inf.
  Theory}, vol.~52, no.~9, pp. 3936--3964, Sep. 2006.

\bibitem{Vishwanath:duality_achievable:2003}
S.~Vishwanath \emph{et~al.}, ``Duality, achievable rates and sum-rate capacity
  of {Gaussian MIMO} broadcast channels,'' \emph{{IEEE} Trans. Inf. Theory},
  vol.~49, no.~10, pp. 2658--2668, Oct. 2003.

\bibitem{Yu:SumCapacity:MIMO_BC:Decomposition:2006}
W.~Yu, ``Sum-capacity computation for the {Gaussian} vector broadcast channel
  via dual decomposition,'' \emph{{IEEE} Trans. Inf. Theory}, vol.~52, no.~2,
  pp. 754 --759, Feb. 2006.

\bibitem{tang2020wireless}
W.~Tang \emph{et~al.}, ``Path loss modeling and measurements for reconfigurable
  intelligent surfaces in the millimeter-wave frequency band,'' \emph{arXiv
  preprint arXiv:2101.08607}, 2021.

\end{thebibliography}
